\documentclass{mn2e}
\usepackage{graphics}
\usepackage{epsfig}
\usepackage{color}
\usepackage{xspace}
\usepackage{times}

\usepackage[latin1]{inputenc}

\def\etal{et\thinspace al.\ }                               

\newcommand{\Ha}{\ifmmode {\rm H}\alpha \else H$\alpha$\fi\xspace}
\newcommand{\Hb}{\ifmmode {\rm H}\beta \else H$\beta$\fi\xspace}
\newcommand{\Hg}{\ifmmode {\rm H}\gamma \else H$\gamma$\fi\xspace}
\newcommand{\Hd}{\ifmmode {\rm H}\delta \else H$\delta$\fi\xspace}

\newcommand{\Hii}{\ifmmode \rm{H}\,\textsc{ii} \else H\,{\sc ii}\fi}
\newcommand{\Nii}{[N\,{\sc ii}]$\lambda$6584}
\newcommand{\nii}{\ifmmode [\rm{N}\,\textsc{ii}] \else [N\,{\sc ii}]\fi}
\newcommand{\Oi}{[O\,{\sc i}]$\lambda$6300}
\newcommand{\oi}{\ifmmode [\rm{O}\,\textsc{i}] \else [O\,{\sc i}]\fi}
\newcommand{\Oii}{[O\,{\sc ii}]$\lambda$3727}
\newcommand{\neiii}{\ifmmode [\rm{Ne}\,\textsc{iii}] \else [Ne\,{\sc iii}]\fi}
\newcommand{\Neiii}{[Ne\,{\sc iii}]$\lambda$3869}
\newcommand{\hei}{\ifmmode [\rm{He}\,\textsc{i}] \else [He\,{\sc i}]\fi}
\newcommand{\oii}{\ifmmode [\rm{O}\,\textsc{ii}] \else [O\,{\sc ii}]\fi}
\newcommand{\Oiii}{[O\,{\sc iii}]$\lambda$5007}

\newcommand{\oiii}{\ifmmode [\rm{O}\,\textsc{iii}] \else [O\,{\sc iii}]\fi}
\newcommand{\Sii}{[S\,{\sc ii}]$\lambda\lambda$6716,6731}
\newcommand{\sii}{\ifmmode [\rm{S}\,\textsc{ii}] \else [S\,{\sc ii}]\fi}
\newcommand{\siii}{\ifmmode [\rm{S}\,\textsc{iii}] \else [S\,{\sc iii}]\fi}

\newcommand{\msun}{M$_\odot$}
\newcommand{\zsun}{Z$_\odot$}

\title[How to distinguish AGN hosts]
          {Semi-empirical analysis of Sloan Digital Sky Survey galaxies\\
          III. How to distinguish AGN hosts}

\author[Stasi\'nska et al.]
{Gra\.zyna Stasi\'nska$^{1}$\thanks{E-mail: grazyna.stasinska@obspm.fr},
Roberto Cid Fernandes$^{2}$, Ab{{\'\i}}lio Mateus$^{3}$, 
Laerte Sodr\'e Jr.$^{3}$,
\newauthor
Natalia V. Asari$^{2}$
\\
$^{1}$LUTH, Observatoire de Meudon, 92195 Meudon Cedex, France\\
$^{2}$Departamento\ de F\'{\i}sica -- CFM -- Universidade Federal de
Santa Catarina, Florian\'opolis, SC, Brazil\\
$^{3}$Departamento de Astronomia, IAG-USP, Rua do Mat\~ao 1226, 05508-090,
S\~ao Paulo, Brazil\\
}
\begin{document}
\pagerange{\pageref{firstpage}--\pageref{lastpage}} \pubyear{2005}

\maketitle

\begin{abstract}
This paper considers the techniques to distinguish normal star forming
(NSF) galaxies and active galactic nuclei (AGN) hosts using optical
spectra. The observational data base is a set of 20\,000 galaxies
extracted from the Sloan Digital Sky Survey, for which we have
determined the emission line intensities after subtracting the stellar
continuum obtained from spectral synthesis. Our analysis is based on
photoionization models computed using the stellar ionizing radiation
predicted by population synthesis codes (essentially Starburst 99)
and, for the AGNs, a broken power-law spectrum. We explain why, among
the four classical emission line diagnostic diagrams, (\oiii/\Hb\ vs
\oii/\Hb, \oiii/\Hb\ vs \nii/\Ha\ (the BPT diagram), \oiii/\Hb\ vs
\sii/\Ha, and \oiii/\Hb\ vs \oi/\Ha), the BPT one works best. We show
however, that none of these diagrams is efficient in detecting
AGNs in metal poor galaxies, should such cases exist. We propose a new
divisory line between ``pure''  NSF galaxies and AGN hosts: $ y =
(-30.787+1.1358x+0.27297x^2){ \rm tanh}(5.7409x) -31.093, $ where $y =
$ log (\oiii/\Hb), and $x$= log (\nii/\Ha).  According to our models,
the divisory line drawn empirically by Kauffmann et al. (2003)
includes among  NSF galaxies objects that may have an AGN contribution
to \Hb\ of up to 3\%.  The Kewley et al. (2001) line allows for an AGN
contribution of roughly 20\%.  About 20\% of the galaxies in our
entire sample that can be represented in the BTP diagram 
 are found between our divisory line and the
Kauffmann et al. line, meaning that the local Universe contains a fair
proportion of galaxies with very low level nuclear activity, in
agreement with the statistics from observations of nuclei of 
nearby galaxies.  We also show that a classification into NSF
and AGN galaxies using only \nii/\Ha is feasible and useful.

Finally, we propose a new classification diagram, the $DEW$ diagram, plotting $D_n(4000)$ vs
max(EW\oii,EW\neiii). This diagram can be used with  optical
spectra for galaxies with redshifts up to $z = 1.3$, meaning an
important progress over classifications proposed up to now.  Since
 the $DEW$ diagram requires only a small range in wavelength, it can also be
used at even larger redshifts in suitable atmospheric windows. It also
has the advantage of not requiring stellar synthesis analysis to
subtract the stars and of allowing one to see \emph{all} the galaxies
in the same diagram, including passive galaxies.
\end{abstract}



\label{firstpage}

\section{Introduction}

Until recently, it was believed that active galactic nuclei 
were found in only a small fraction of all galaxies (Huchra \& Burg 
1992). However, it was already known that a large fraction of 
galaxies have nuclei with a very low level activity (called LINERs by 
Heckman 1980, for Low Ionization Nuclear Emission Regions), and that 
this activity would not be detectable in distant galaxies.

Generally, normal star forming (NSF) galaxies are distinguished from
those containing an active galactic nucleus (AGN) using diagrams where
are plotted emission line ratios.  The most common diagnostic diagrams
are those of Baldwin, Phillips \& Terlevich (1981, BPT) and Veilleux
and Osterbrock (1987, VO). The lines in  NSF galaxies
are emitted by HII regions, which are ionized by massive stars, while
AGNs are ionized by a harder radiation field. Therefore, for a given
\oiii/\Hb\ or \oiii/\oii\ ratio, AGN galaxies will show higher
\oii/\Hb, \nii/\Ha, \sii/\Ha, or \oi/\Ha\ ratios \footnote {In the entire
paper  \oiii\ stands for \Oiii, \oii\ for \Oii,
\nii\ for \Nii, \sii\ for \Sii, and \oi\ for \Oi.}, than NSF
galaxies. The dividing line between NSF galaxies and AGN hosts has
slightly changed over the years. In BPT and VO, it was a compromise
between what was suggested by a limited number of data points and some
coarse grids of crude photoionization models. More recently,
Kewley et al. (2001) proposed a theoretical boundary,  defined by the
upper envelope of their grid of photoionization models in which the
ionizing source was provided by young stellar clusters. 

With the advent of the Sloan Digital Sky Survey (SDSS, York et
al. 2000, Abazajian et al. 2004), the number of data points increased
by orders of magnitude.  Also, techniques to model the stellar
component of the spectra and subtract it from the observed spectrum to
obtain the pure nebular spectrum became practicable on a large number
of objects. As a result, in the \oiii/\Hb vs. \nii/\Ha\ diagram, the
$\sim$ 50000 SDSS galaxies having S/N $>3$ in all the four lines and
pertaining to a complete sample of about 120000 galaxies clearly
outline two wings (Kauffmann et al. 2003) which look like the wings of
a seagull.  
Kauffmann et al. (2003) have defined a purely
empirical dividing line between NSF and AGN galaxies. This dividing
line is significantly below the line drawn by Kewley et al. (2001).

Interestingly, the SDSS has definitely shown that, in the local
Universe, the number of galaxies hosting AGNs is of the same order as
that of NSF galaxies (within a factor which depends on selection
criteria and definitions). Studies based on other galaxy samples
(e.g. Carter et al.  2001) also came to a similar conclusion, but the
SDSS results are stronger, being based on a much larger number of
objects, a clear selection function, high resolution spectra and
elaborate subtraction of stellar features.

There is actually an important difference between the original BPT or
VO diagrams and the Kauffmann et al. (2003) diagram. The former were
constructed using spectra of known giant HII regions (mainly located
in spiral galaxies) and known  nearby active galactic nuclei,
while the Kauffmann et al. (2003) plot concerns galaxy spectra
obtained  through  3\arcsec\ fibres  which, at their $z \sim
0.1$ typical redshift, corresponds to 6 kpc 
(for $H_0=70$ km s$^{-1}$ Mpc$^{-1}$) . Hence, in many galaxies, the
region covered by the fibre encompasses a significant fraction of volume and light of the entire galaxy.  Thus, galaxies that occupy the same position as
LINERs in these diagrams a priori have no reason to be galaxies
\emph{hosting} a LINER, since the emission line flux from the low
ionization nuclear emission region is small with respect to the
emission line flux from a  region of several kiloparsecs in diameter, at least in galaxies which still form stars.

One of the persistent questions in  astronomy  is what causes or favors 
non-stellar activity in galaxies (see e.g. the proceedings of  the 
IAU symposium ``The Interplay among Black Holes, Stars and ISM in 
Galactic Nuclei'', Storchi-Bergmann et al. 2004). The SDSS is 
revolutionizing our ways to attack this problem (e.g. Heckman et al. 
2004, Kauffmann et al. 2004, Best et al., 2005, Fukugita et al. 2004, 
Hao et al. 2005a,b, Pasquali et al. 2005), and deeper surveys will 
follow. In view of this, it is important to revisit the classification 
criteria of galaxies in order to lay them on sounder ground. This is 
the purpose of the present paper.

The paper is organized as follows. In Sect. 2, we present the data 
sample, and the method to measure emission line intensities. In Sect. 
3 we show and discuss some classical emission line diagrams. In Sect. 4 we 
compare the distribution of observational points with the location of 
photoionization models for giant HII regions.  In  Sect. 5, we 
propose a simple model to  account for the emission line properties of 
AGN host galaxies. In Section 6, we present our boundaries to 
distinguish NSF galaxies and AGN hosts in classical emission-line diagrams, and
we propose alternative classifications, including one that can be easily used for high redshift objects. The last section 
summarizes our results.

\section{The data}

\subsection{The sample}

 The data used in this work were taken from the SDSS. The most relevant
characteristic of this survey for our study is the enormous amount of good 
quality, homogeneously obtained spectra. We consider a flux-limited sample 
extracted from the SDSS main galaxy sample available in the Data Release 2 
(Azabajian et al. 2004). From such database we have selected at random 
20\,000 galaxies with reddening-corrected Petrosian $r$-band magnitudes $r 
\le 17.77$, and Petrosian $r$-band half-light surface brightnesses $\mu_{50} 
\le 24.5$ mag arcsec$^{-2}$ (Strauss \etal 2002). As a quality cut, we 
restricted our sample to objects for which the observed spectra show a $S/N$ 
ratio in $g$, $r$ and $i$ bands greater than 5. The median value of redshift 
for this sample is $z = 0.097$ and the galaxies have a median $r$-band 
absolute magnitude of $M_r = -21.41$. We note that Seyfert~1 objects are not 
included in our sample.

\subsection{The spectral synthesis of the stellar continuum}

The SDSS spectra cover a wavelength range of 3800--9200 \AA, have mean
spectral resolution $\lambda/\Delta\lambda \sim 1800$, and were taken
with 3\arcsec\ diameter fibres. The spectra are first corrected for
Galactic extinction using the maps of Schlegel, Finkbeiner \& Davis
(1998) and using the extinction law of Cardelli et al. (1989). 
They are then brought to the rest-frame and resampled from 3400 to
8900  \AA\ in steps of 1 \AA~ with a flux normalization by the median
flux in the 4010--4060 \AA~region.

To measure the intensities of the emission lines, we   have to 
subtract the stellar continuum. This is done by computing for each 
galaxy a synthetic stellar spectrum which is a combination of simple 
stellar population (SSP) spectra and fits the observed continuum in 
the entire spectral range (after removal of the zones of emission 
lines and bad pixels). The method, implemented in the STARLIGHT code, 
is fully described in Cid Fernandes et al. (2005, hereafter SEAGal I) and 
Mateus et al. (2006, SEAGal II). As in SEAGal II, we use a base of 
150  SSPs, spanning 6 metallicities: $Z=0.005, 0.02, 0.2, 0.4, 1$ and 
2.5 $Z_\odot$, with  25 different ages between 1 Myr and 18 Gyr. 
Extinction by dust in the galaxy is taken into account in the 
synthesis, assuming that it arises from a  foreground screen  with the
extinction law of Cardelli et al. (1989). In SEAGal I, we have shown 
that this simple method is capable of reproducing the stellar 
continua of real galaxy spectra  very  well. It therefore provides a 
reliable estimate of the stellar absorption in the entire spectral 
range, including the windows where emission lines are found. For each 
galaxy, we thus obtain the pure emission line spectrum by 
subtracting the synthetised stellar spectrum from the observed one.

\subsection{Emission line measurements and dereddening}

We have developed a code to measure the main emission
lines from the pure emission line spectrum by fitting them 
as Gaussian functions, composed by three parameters:
width, offset (with respect to the rest-frame central wavelength), and flux.
Lines from the same ion are assumed to have the same width and offset.
Additionally,  we consider the following flux ratio constraints: 
\oiii$\lambda5007$/\oiii$\lambda4959 = 2.97$ and
\nii$\lambda6584$/\nii$\lambda6548 = 3$.
The currently measured lines following this approach include:
\oii$\lambda\lambda$3726,3729, H$\delta$, H$\gamma$, H$\beta$,
\oiii$\lambda\lambda$4959,5007, \oi$\lambda$6300, \nii$\lambda$6548,
H$\alpha$, \nii$\lambda$6584 and \sii$\lambda\lambda$6716,6731, among
others that we will include as our needs increase.  In this way, for
each emission line, our code returns the rest-frame flux and its
associated equivalent width (EW), the velocity dispersion measured
from the line width, the velocity displacement relative to the
rest-frame wavelength, and the $S/N$ of the fit. Note that, with our 
approach, the Balmer lines EWs are not affected by the underlying stellar
absorption.  The emission line ratios have been dereddened using the
standard Cardelli et al. (1989) extinction law ($R_V$=3.1) and
adopting an intrinsic \Ha/\Hb\ of 2.86 for all the galaxies. As a
matter of fact, this correction is unimportant for all the line ratios
we consider in this study, except for \Oii/\Hb.   One might worry whether neglecting the dependence of the intrinsic  \Ha/\Hb\ ratio with metallicity induces a sizeable bias. We find  the bias to be less than 40\% for  \Oii/\Hb\ between the most metal-poor and the most metal-rich objects in the sample. For the other line ratios  considered in this paper, the bias is completely negligible.

\subsection{Comparison of our  data analysis techniques with those reported in the literature}

 Both our starlight subtraction and emission line measuring procedures
are very similar to those followed by Tremonti et al. (2004) and
employed in a series of papers (eg, Kauffmann et al. 2003). The
differences are merely technical, like which SSPs are included in the
starlight modelling and which constraints are applied when fitting the
emission lines. As shown in Cid Fernandes et al. (2005),
our emission line fluxes and equivalent widths are in excellent
agreement with those published by Brinchmann et al. (2004).

\begin{figure*}
\centerline{
\includegraphics[scale=0.6]{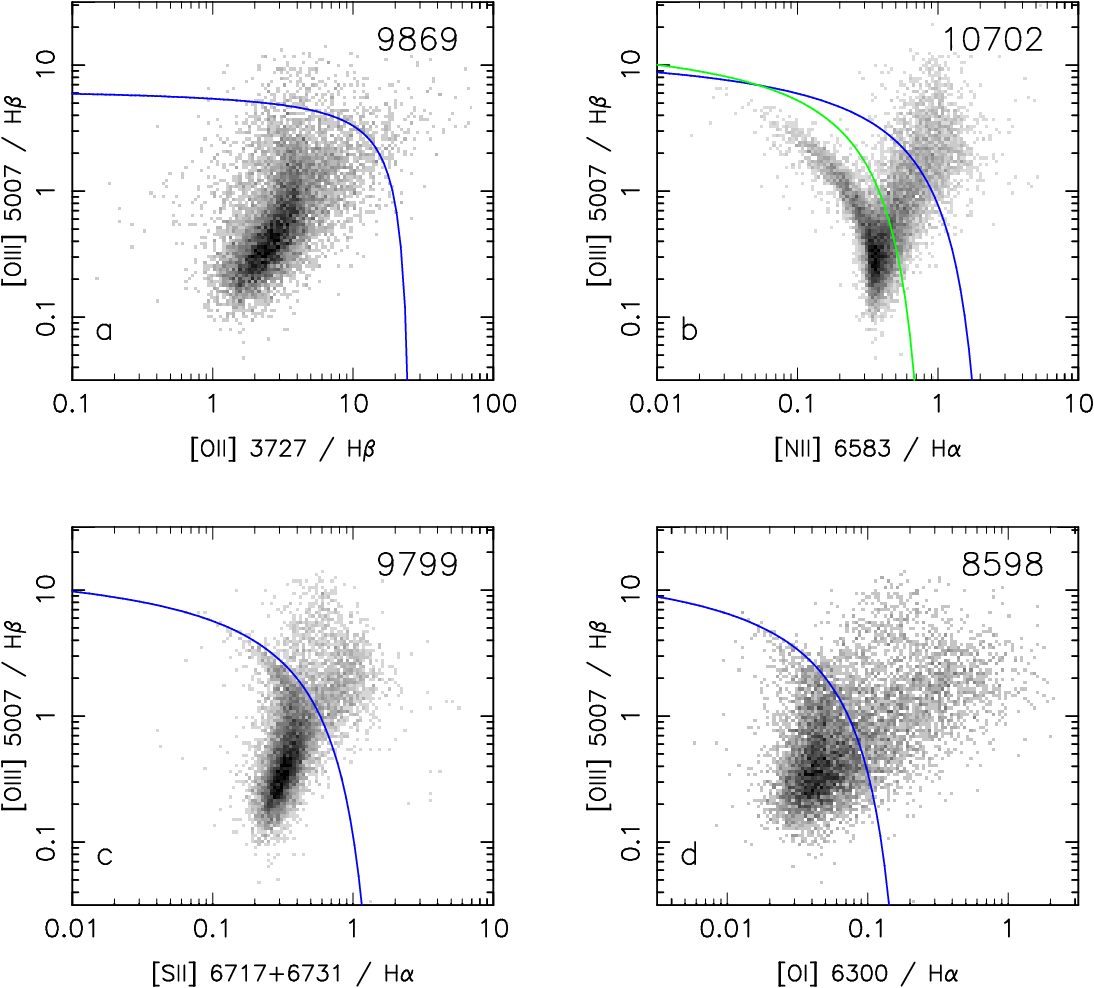}
}
\caption{Our sample galaxies in four classical emission line ratio 
diagnostic diagram: \oiii/\Hb\ vs \oii/\Hb\ (a), \oiii/\Hb\ vs 
\nii/\Ha\ (b), \oiii/\Hb\ vs \sii/\Ha\ (c) ,  \oiii/\Hb\ vs \oi/\Ha\ 
(d).  The grey scale level represents the number of galaxies in each 
pixel, darker pixels being more populated. The total number of 
galaxies in the various plots is indicated in the top right of each 
panel. In panel $a$, the line (blue in the on-line version of the paper) represents the empirical curve of Lamareille et al. (2004). In panel $b$, $c$, and $d$ the blue lines represent the Kewley et al. (2001) lines. The green line in panel $b$ represents the  
Kauffmann et al. (2003) line}.
\label{1}
\end{figure*}


\section{Classical emission line diagrams}

Figure 1 shows four classical diagnostic diagrams used to distinguish
NSF galaxies from galaxies containing an active nucleus. These are
diagrams based on line intensity ratios. Three of them have been
popularized by Veilleux \& Osterbrock (1987) and have been widely used
since then:  \oiii/\Hb\ vs \nii/\Ha\ (panel b), \oiii/\Hb\ vs \sii/\Ha\ (panel c),  and  \oiii/\Hb\ vs \oi/\Ha\ (panel d). The fourth one,
\oiii/\Hb\ vs \oii/\Hb\ (panel a), has been used e.g. by Tresse et
al. (1995) or Lamareille et al. (2004).   The \oiii/\Hb\ vs
\nii/\Ha\ has actually been introduced by Baldwin et al. (1981), and
will be referred to as the BPT diagram. The total number of galaxies
in each diagram is indicated in the plots. In order to do these plots,
we have imposed no condition on the signal-to-noise in the line, thus
an object appears here as soon as we are able to measure the
intensities of all the emission lines involved in the plot.  Thus,
about half of the galaxies of our initial sample have at least four relevant
emission lines detected and are represented in this diagram. The rest of the
galaxies have either between 1 and 3 of those lines detected, or none of
them. Galaxies of this latter group are called passive galaxies.
 Note that restricting the  diagrams of Fig. 1 only to
objects with a signal-to-noise ratio of at least 3 in each relevant
line does not change the apparent distribution of points in the plots,
but reduces the proportion of objects in the right wing and in the
``body'' of the seagull.

The basic idea underlying these diagrams is that the emission lines
in NSF galaxies are powered by massive stars, so that there is a well 
defined upper limit on the intensities of collisionally excited lines 
with respect to recombination lines (such as \Ha\ or \Hb). In 
contrast, AGNs are powered by a source of much more energetic photons 
so that, globally, collisionally excited lines are more intense, 
implying that galaxies hosting AGNs should be found to the upper 
right of NSF galaxies in these diagrams.  It has long been known that 
giant HII regions actually form a very narrow sequence in these 
diagrams (see eg. Mc Call et al. 1985). This implies that, while a 
priori the emission line ratios of giant HII regions are defined by 
three main parameters (namely the metallicity, the mean effective 
temperature of the ionizing stars and the ionization parameter), 
these three parameters must be linked together and one may say that 
the observed sequence is essentially driven by metallicity. The 
physical reason  behind the HII region sequence is not yet clear, but 
this is an observational fact.
Recent spectroscopic surveys of galaxies (e.g. Jansen et al. 2000, 
Moustakas \& Kennicutt 2006) have shown that the emission line 
sequence of NSF galaxies is actually very close to the giant HII 
region sequence. The SDSS, with its thousands of galaxies, shows a 
superb, very narrow sequence in the BPT diagram 
(the extension to the upper left is very faint in  Fig. 1b, because 
only a small fraction of our sample populates this region of the 
diagram, which corresponds to very low metallicity star forming galaxies). 

The big surprise, with the SDSS, was the apparition of a second
sequence, starting from the bottom of the HII region sequence and
extending to the upper right of the diagram. This sequence is fuzzier
than the HII region sequence, but nonetheless  clearly present. Thus, this
suggests that line emission in AGN galaxies is shaped by one
dominant parameter or by a set of correlated parameters.  As a matter of fact, this trend was already suggested in the
sample of 285 warm IRAS galaxies studied  by Kewley et al. (2001), but it
became conspicuous only with SDSS data.  Interestingly, the sequence
of AGN host galaxies (the right wing of the seagull) appears clearly
only in the \oiii/\Hb\ vs \nii/\Ha\ diagram. As seen in Fig. 1, it is
very ``fuzzy'' and forms a small angle with the HII region sequence
in the \oiii/\Hb\ vs \sii/\Ha\ and \oiii/\Hb\ vs \oi/\Ha\ diagrams and
almost merges with the HII region sequence in the \oiii/\Hb\ vs
\oii/\Hb diagram.

The existence of this right wing is obviously of extreme importance
for our understanding of the AGN phenomenon. It has been analyzed
empirically by Kauffmann et al. (2003) and shown to be linked to the
\oiii\ luminosity and to the mass of the parent galaxy. 

\section{Normal star forming galaxies: the left wing of the seagull}

\subsection{Preliminaries}

In studies dealing with statistics of the AGN phenomenon, it is
important to have a clear criterion to detect the presence of an
AGN. Dopita et al. (2000) and Kewley et al. (2001) have constructed an
extensive grid of photoionization models for giant HII regions powered
by star clusters.  The ionizing radiation field is provided by stellar
synthesis models assuming two limiting cases: a constant star formation rate and an instantaneous starburst. They have used different population synthesis codes (PEGASE 2: Fioc \& Rocca Volmerange 1997 and Starburst99: Leitherer et al. 1999) and with each code experimented all the available stellar evolutionary track and  atmosphere sets. The models
were defined by two parameters: the metallicity and the ionization
parameter.  Kewley et al. (2001) used the entire set of  models to define an upper
envelope in the diagnostic diagrams of regions that can be powered
only by HII regions. As noted by Kauffmann et al.  (2003), this upper
envelope -- the ``Kewley et al. line'' -- is actually well above the
NSF sequence delineated by SDSS galaxies. 

Two comments are in order. One is that the models of
Dopita et al. (2000) were built just
before the latest model atmospheres for massive stars (Pauldrach et
al. 2001 and Hillier \& Miller 1998)  were incorporated in public
stellar population synthesis codes. These models, which include the
effect of non-LTE, mass loss and line blanketing, have a softer
radiation field at high metallicity than previous models.  Second,
there is a priori no reason why the upper envelope should correspond to
the observed NSF galaxy sequence.  This last argument prompted Kauffmann et
al. (2003) to draw an \emph {empirical} curve separating NSF galaxies
from AGN hosts in the \oiii/\Hb\ vs \nii/\Ha\ diagram. It is not quite
clear from their paper how they defined this curve, and it too lies
slightly above the NSF galaxy sequence (especially in the upper left
of the diagram). In any case, it is by extrapolation that they defined
the curve in the zone of low values of \oiii/\Hb, where the two wings
of the seagull come into contact with its body.

The  \oiii/\Hb\ vs \nii/\Ha\ diagram is the most commonly used  to separate NSF galaxies from AGN hosts (see e.g. Brinchman et al. 2004, Lamareille et al. 2004, Mouhcine et al. 2005, 
Gu et al. 2006). In this section, we will look for a sequence of models that  fits the upper envelope of the NSF galaxy sequence, and see how this sequence translates in the other traditional diagnostic diagrams shown in Fig. 1. 

\subsection{Our starting photoionization model grid} 

Taking advantage of the implementation by Smith et al. (2002) of the
Pauldrach et al. (2001) and Hillier \& Miller (1998) stellar
atmospheres into the Starburst99 code of Leitherer et al. (1999), we
have run a grid of photoionization models using the spectral energy
distribution provided by that code feeded into the photoionization code PHOTO
(using the version described in Stasi\'nska 2005). We have used
standard constant star formation models, which are the most
appropriate for galaxies containing a large number of HII regions of different ages.  We
adopted a Salpeter IMF and an upper stellar mass limit of 120\msun,
which is the canonical parameterization for such kinds of studies. We
took the Starburst99 option that uses the Geneva tracks with high mass
loss.  As explained by V\'azquez \& Leitherer (2005), this is the recommended option when interested in ionizing spectra. The models of our grid are computed for the following
metallicities: $Z$ = 0.1, 0.2, 0.3, 0.4, 0.6, 0.8, 1.0, 1.5 and
2.5\,\zsun. The metallicities are the same for the nebular gas and for
the stars. For the stars, we interpolated from the spectral energy distributions at the two bracketing metallicities available in Starburst99.  For the nebulae, we considered that the
metallicity is defined by the oxygen abundance, taking a solar value
of O/H = $4.9 \times 10^{-4}$ (Allende Prieto et al. 2001). We adopted the
solar He/H ratio of Grevesse \& Sauval (1998) for all the models.  The
abundances of the $\alpha$-elements with respect to oxygen were chosen
to follow the laws found empirically by Izotov et al. (2006) for
metal-poor emission line galaxies. For the most important elements in
our context, we thus have:
\begin{equation}
\log \frac{\rm Ne}{\rm O} = 0 .088{\rm X} - 1.450,
\end{equation}
\begin{equation}
\log \frac{\rm S}{\rm O} = -0.026{\rm X} - 1.514,
\end{equation}
where X = 12 + log O/H.
For nitrogen, we also based on the paper by Izotov et al. (2006), 
using their Fig. 11 to adopt the following law:
\begin{equation}
\log \frac{\rm N}{\rm O} = -1.6   {,\,\,\,\,\,\,\,\,  \rm    for ~  X} < 8,
\end{equation}
and
\begin{equation}
\log \frac{\rm N}{\rm O} = 0.6({\rm X}-8) - 1.6 {,\,\,\,\,\,\,\,\, 
\rm      for ~ X} > 8
\end{equation}
Note that these abundance ratios are different from the solar ones as 
compiled by Lodders (2003). Since we are interested in emission
lines in galaxies, it is more natural to use abundance ratios that 
are indicated by abundance analysis in such galaxies. \footnote {As a matter of
fact, with the solar abundance ratios, our photoionization models do not reproduce  the distribution of SDSS galaxies in all the line ratio diagrams simultaneously.}

The models are computed for thin bubbles with a constant hydrogen density of
$n_H$ = 100 cm$^{-3} $ (note that, within the range of densities typical of the giant HII regions in emission line galaxies, our results are not affected by the choice of $n_H$). The models are characterized by an ionization parameter
$U$, defined as $U=Q_H/(4 \pi R^2 n_H c)$, where $Q_H$ is the total
number of H-ionizing photons emitted per second by the stars, $R$ is the radius
of the bubble in cm and $c$ is the speed of light. The chosen values
of $U$ are 10$^{-2}$, $5 \times 10^{-3}$, $2 \times 10^{-3}$, 10$^{-3}$, 
$5 \times 10^{-4}$, and $2 \times 10^{-4}$. For each value of $U$, we compute a sequence with
varying abundances, as explained above.  This is obviously a very
crude way to model the spectrum of a galaxy seen through a 3\arcsec\
fibre, as it accounts neither for abundance gradients, nor for
diffuse emission, nor for the complex structure of realistic HII
regions. Still, they provide useful guidelines to interpret the
observed data.

\begin{figure*}
\centerline{\includegraphics[scale=0.6]
{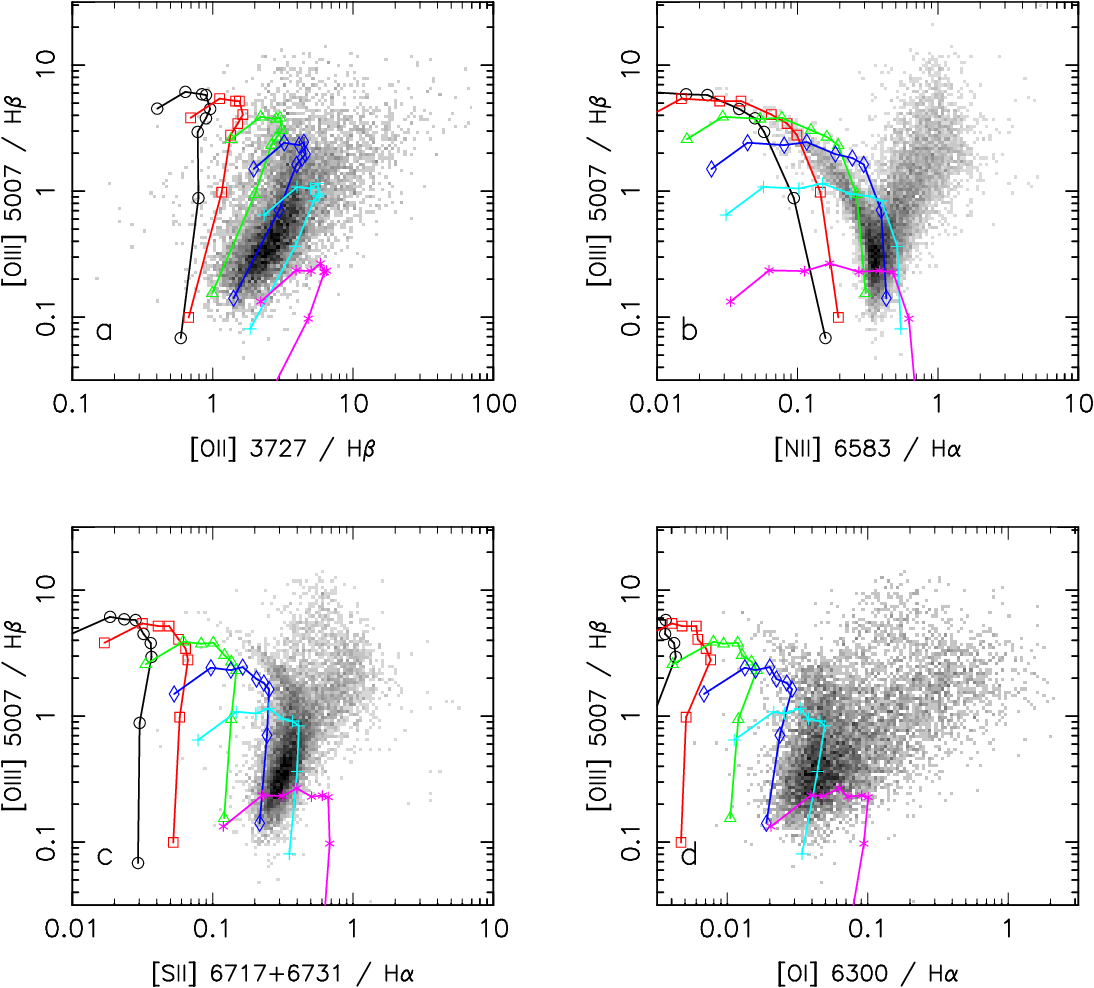}}
\caption{Sequences of photoionization models with varying metallicities $Z$  and constant ionization parameter $U$. The symbols on the curves correspond to the location of models with metallicities $Z$ = 0.1, 0.2, 0.3, 0.4, 0.6, 0.8, 1.0, 1.5 and
2.5\,\zsun, going from the upper left to the lower right  (in panels b, c and d,  the lowest metallicity models are actually outside the range of the plots).
The values of the ionization parameter $U$ are 10$^{-2}$ (black
circles), $5 \times 10^{-3}$ (red squares), $2 \times 10^{-3}$ (green
triangles),10$^{-3}$ (blue diamonds), $5 \times 10^{-4}$ (cyan + signs),
$2 \times 10^{-4}$ (purple * signs).}
\label{bl}
\end{figure*}

These model sequences are shown in Fig. 2. We can see that the upper 
limit of our sequences lies rather close to the left wing of the 
seagull in the  BPT diagram, especially in the 
upper part. It is well below the Kewley et al. line (in blue in Fig. 1b, 
and even below the Kauffmann et al. line (in green in Fig. 1b). As noted in former studies (e.g. Dopita et al. 
2000), the observed sequence of star forming galaxies corresponds to 
only a small selection of a grid sampling a whole range of values of $U$ and 
$Z$.

Note that the model sequences have slightly different shapes in the 
various diagrams of Fig. 2. In the \oiii/\Hb\ vs \oii/\Hb\ diagram, as 
the metallicity increases from  $Z$ $\simeq$  0.8\,\zsun\ onwards, 
models with same $U$ move down and slightly to the left. This is due 
to the well-known fact that, with increasing cooling, the electron 
temperature drops, and lines that require a significant amount of 
energy to be excited, such as \oiii\ or \oii, become weaker. Why then 
do the sequences of constant $U$ drop rather vertically in the 
\oiii/\Hb\ vs \sii/\Ha\  and \oiii/\Hb\ vs \oi/\Ha\ diagrams? The 
reason is that the \sii\ and \oi\ lines require less energy than 
\oii\ to be excited, so that the drop in electron temperature is 
compensated by the increase in element abundance and the intensity of 
these lines remains roughly constant. Although the \nii\ line has an 
excitation potential intermediate between that of \oi\ and that of 
\sii, the lines of constant $U$ in the \oiii/\Hb\ vs \nii/\Ha\  fall down 
towards the right, i.e. \nii/\Ha\ {\em increases}. This is because, in 
our models, N/O increases with O/H at large metallicity. It is this 
different behaviour of the low excitation lines that leads to the 
different aspects of the observational plots in the various panels of 
Figs. 1 and 2.

\subsection{A sequence of models for the upper envelope of the NSF galaxy sequence}

We can use our grid to look for an empirical relation between the
ionization parameter and the metallicity that will satisfactorily
delineate NSF galaxies in the BPT diagram.

\begin{figure*}
\centerline{\includegraphics[scale=0.6]
{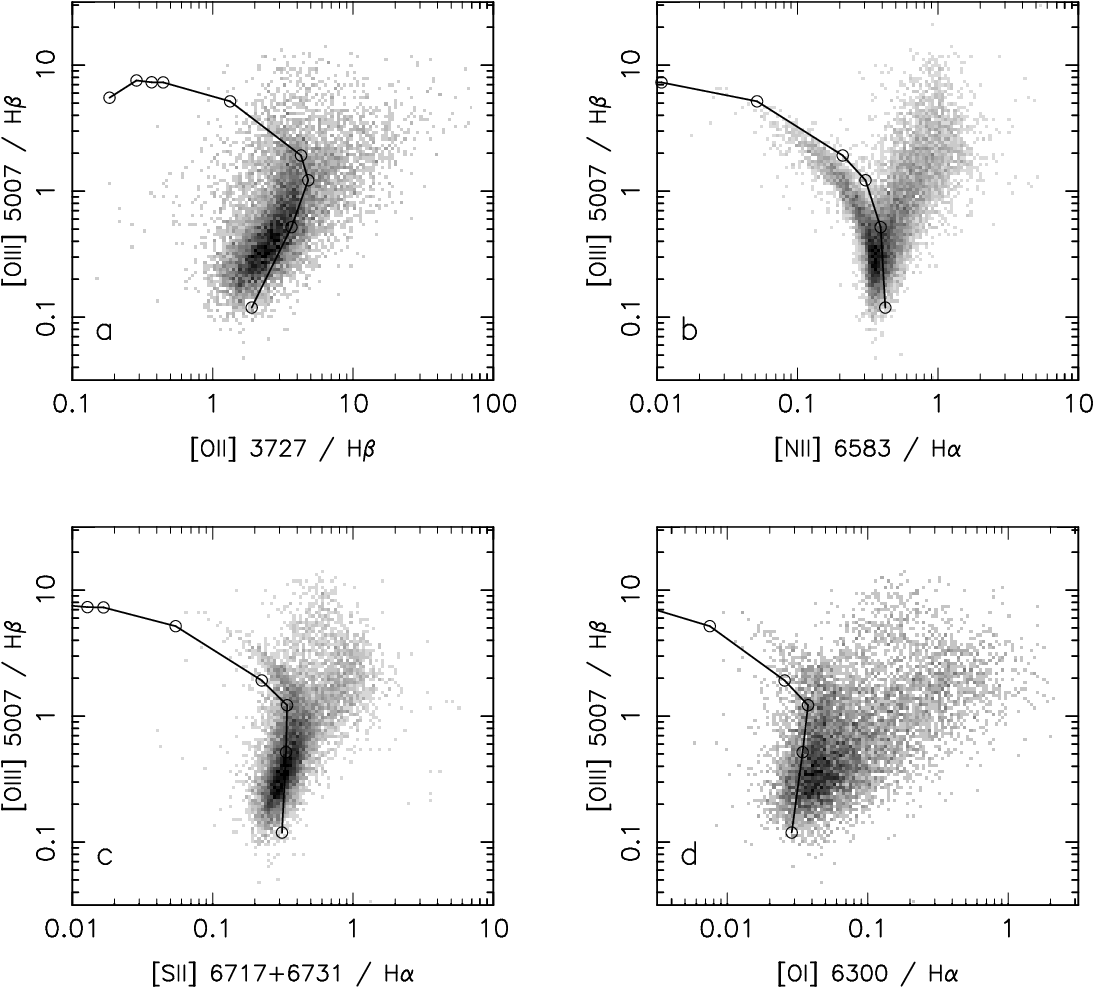}}
\caption{The sequence of photoionization models  
defined by $ \log U = 0.916 \tanh (-5.42\times Z+3.28)-2.26 $ (see Sect. 
4.3), and the data from our SDSS sample galaxies, in the four classical diagnostic diagrams.}
\label{fig_aperture_effects}
\end{figure*}

With the stellar radiation field used in our starting models, we had
difficulties in reproducing the tip of the left seagull wing: all the
models had slightly too low \oiii/\Hb, meaning that the radiation
field is not hard enough. Since the tip of the left wing corresponds
to low mass and metallicity galaxies, it is likely that, in most of
them, the ionizing radiation field is dominated by that from a recent
starburst, as opposed to more massive and metal-rich galaxies which
populate the bottom of the left wing, and have a more continuous
star-formation regime (Cid Fernandes, Le\~ao \& Lacerda 2003). The
radiation field from a recent star burst is harder than that provided
by stars that are constantly forming at the same rate, because it is
dominated by the most massive stars. In models with metallicities
lower than 0.7\,\zsun, we have then replaced the stellar energy
distribution resulting from a constant star formation rate with that
produced by an instantaneous burst. This improves the modeling of the
tip of the left wing, but still does not make it perfect. We have
tried other options (changing the geometry, using another stellar
synthesis code or another photoionization code),  but the problem
still remains. As a matter of fact, this problem is not new
(Stasi\'nska \& Izotov 2003). Whether it requires an additional
heating source or a more complex modeling of the HII region to be
solved is not yet clear.  Anyway, its consequences on the present
study are minor, so we set it aside from now on.

We find that the upper envelope of the left wing in the BTP diagram is well reproduced by a sequence of 
models in which $U$ and $Z$ are related by:

\begin{equation}
\log U = 0.916 \tanh (-5.42\times Z+3.28)-2.26 .
\end{equation}
Note that, with our description of the radiation field and with the
geometry adopted for the nebular models, there is in principle only one solution for the   $U$ -- $Z$ relation (however, this relation is ill-determined at low $Z$, as can be understood from Fig. 2b).  Other geometries will, of course, lead to 
slightly different relations.

Fig. 3a--d shows the sequence of models defined by Eq. (5) superimposed on the same observational data as in Fig. 1a--d. We note that the sequence defined by Eq. (5) works  also for  the \oiii/\Hb\ vs.\oii/\Hb\ diagram. On the other hand, the models seem to underpredict the values of \sii/\Ha, and even more of \oi/\Ha.  This is not really surprising. It is notorious that simple HII region models produce too small \oi/\Ha\ with respect to observed values in giant HII regions (see e. g. Stasi\'nska \& Leitherer 1996). Therefore, we cannot expect, with our schematic models, to reproduce perfectly the line ratios of SDSS galaxies in all the four diagrams at the same time.  As shown by Stasi\'nska \& Schaerer (1999), explaining \oi\ and \sii\  lines together with the rest of the lines in the spectrum of a giant HII region can be achieved when combining models with different densities. This is  feasible when analyzing an individual object, but not tractable in a study like the present one. We have taken the most reasonable option, i.e. to put emphasis on the \oiii/\Hb\ vs.\oii/\Hb\  and \oiii/\Hb\ vs.\nii/\Ha\ diagrams.

The results from the model sequence defined by Eq. (5) can be parameterized as follows: 
\begin{eqnarray}
\log (\oiii/\Hb) = (0.023606-0.667627 Z)\times \nonumber\\
 \tanh(-3.412213+5.743451 Z)+0.712143
 \end{eqnarray}
\begin{eqnarray}
\log (\oii/\Hb) =  (-0.86928+0.052482 Z)\times \nonumber\\
 \tanh(2.66503+4.4425 Z)-1.2516
\end{eqnarray}
\begin{eqnarray}
\log (\nii/\Ha) =  (-1.0577-0.055221 Z)\times \nonumber\\
 \tanh(2.00404-3.82832 Z)-1.55079
\end{eqnarray}
\begin{eqnarray}
\log (\oi/\Ha) = (-0.83751+0.110241 Z)\times \nonumber\\
 \tanh(2.35279-3.97006 Z)-2.11304
\end{eqnarray}
\begin{eqnarray}
\log (\sii/\Ha) =  (-0.86928+0.052481 Z)\times \nonumber\\
 \tanh(2.66503-4.44255 Z)-1.251617,
\end{eqnarray}

\noindent 
where $Z$ is the model metallicity, with respect to the solar
one. Classification boundaries based on these equations are discussed
in Sect. 6.1. Note that our series of models  for the upper envelope  of the NSF galaxy sequence  lies  much to the left of the Kewley et al. (2001) lines (compare Figs. 3 and  1).  The main reason is that Kewley et al. aimed at producing an ``extreme theoretical starburst line'', which they obtained  using population synthesis models based on Padova evolutionary tracks (Bressan et al. 1993) and  stellar atmospheres that were available at that time. As discussed by V\'azquez \& Leitherer (2005), such population synthesis models  strongly overestimate the hardness of the ionizing radiation field. Another point is that, it is only with the SDSS data that the NSF galaxy sequence became conspicuous, and our series of models  is meant to model its upper envelope -- and not a theoretical upper limit for stellar photoionization.

\section{Modeling AGN hosts}

\begin{figure*}
\centerline{\includegraphics[scale=0.6]
{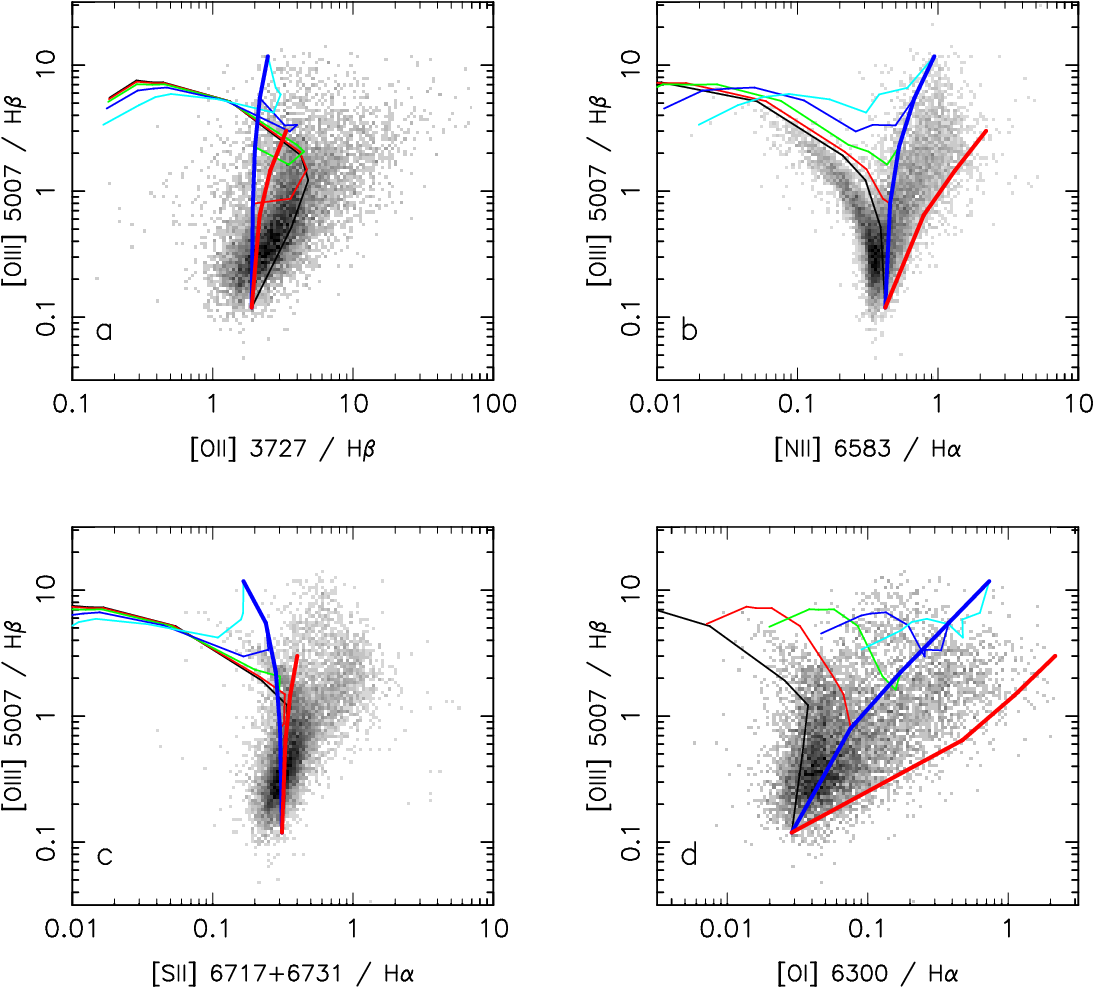}}
\caption{Composite HII regions and AGN models computed by combining 
the intensities obtained in the HII region sequence represented in 
Fig. 3 with those of the AGN model sequence with $U$=0.01 (See Sect. 5.1).
The thin lines correspond to various proportions $\eta$  
between the \Hb\ luminosity from the AGN and from the HII regions:  
$\eta$ = 0 (black), 0.03 (red ), 0.1 (green), 0.3 (blue), 1 
(cyan).  The model metallicities are the same as
in Fig. 2. The thick lines connect the composite models at $Z$ = 2.5\,\zsun\
that have $U$ = 0.01 (blue line) and $U$  = 0.03 (red line). }
\label{mm}
\end{figure*}

\subsection{Composite models}

Let us now turn to AGN hosts. The most common current understanding of
narrow-line AGNs is that the emission lines are due to moderate
density gas ($10^3$ -- $10^5$ cm$^{-3}$) photoionized by a radiation
field extending to the keV region. We thus consider a very simple
model for an AGN, based on the model of Kraemer \& Crenshaw (2000) for
the narrow-line region of the Seyfert galaxy NGC 1068, which is a
broken power-law.  We use a density of $10^4$ hydrogen particles per
cm$^{3}$ and the same radiation field as proposed by Kraemer \&
Crenshaw (2000), and we construct a sequence of photoionization models
having a given ionization parameter and the same abundances as the HII
region model sequences described above.

Inspired by the shape of the BPT diagram, we produced sequences
of composite models for AGN galaxies by adding this AGN model sequence
to the HII region sequence defined by Eq. (5). The results are shown
in Fig. 4, where we plot composite model sequences corresponding to an
ionization parameter $U$ equal 0.01 for the AGN and different values
of the ratio $\eta$ between the \Hb\ luminosity produced by the AGN
and the \Hb\ luminosity produced by the HII regions. The values of
$\eta$ used in Fig. 4 are 0.03 (red curve), 0.1 (green), 0.3 (blue),
and 1 (cyan). The black line represents the models  for the upper
envelope of the pure HII
region sequence defined by Eq. (5). In order to illustrate how
composite models in the right wing depend on the value of $U$ adopted
for the AGN, we also draw two thick lines that connect composite
models with metallicity $Z = 2.5$ \zsun\ for $U$ = 0.01 (blue line) and
$U$ = 0.03 (red line).

\begin{figure*}
\centerline{\includegraphics[scale=0.6]
{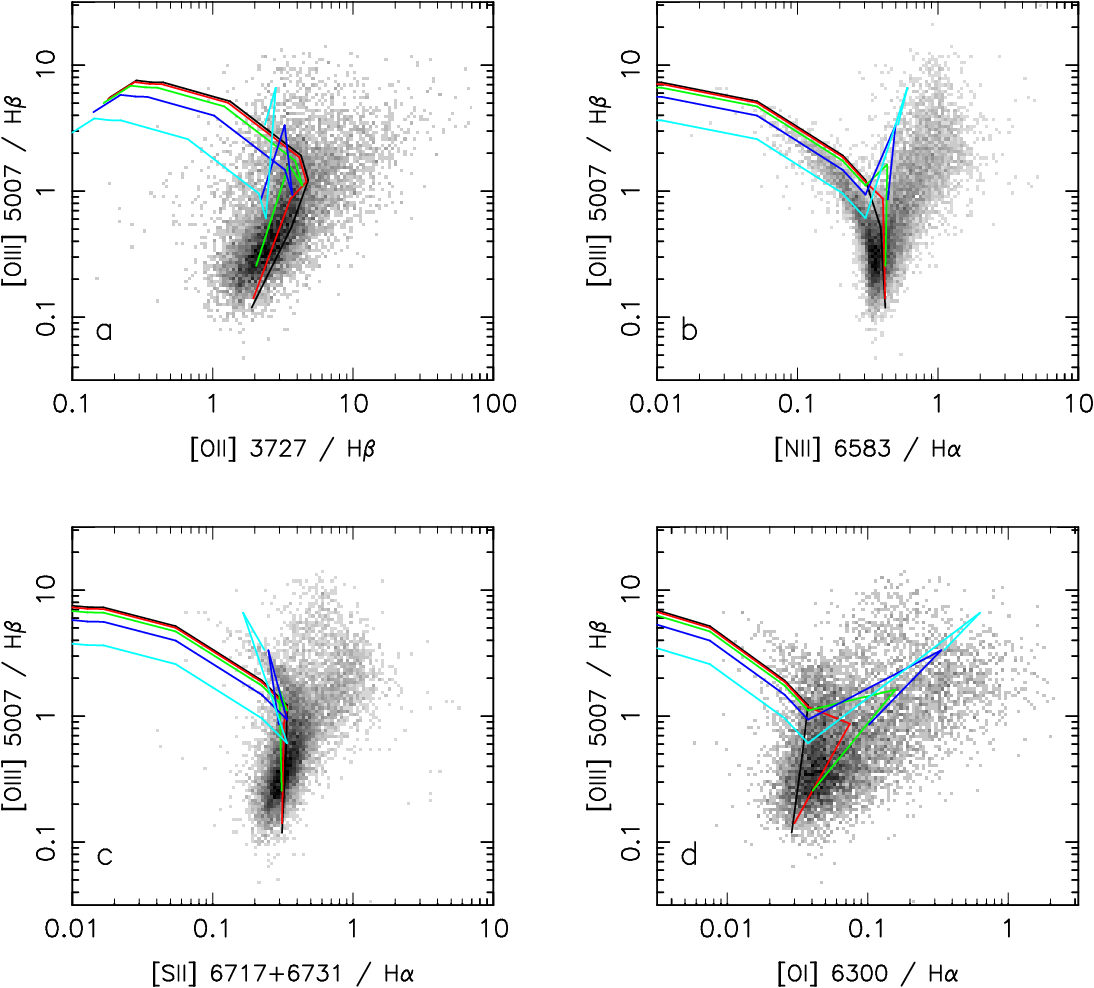}}
\caption{Same as Fig. 4 except that here all the HII region models of
various metallicities from the sequence shown in Fig. 3 are now
combined with the same AGN model of metallicity $Z$ = 1.5\,\zsun\
(and have $U$ = 0.01) }
\label{u3}
\end{figure*}

Fig. 4 shows that such sequences of composite models are very
successful -- given the simplicity of the approach -- in reproducing
the observed trends in the observed diagrams. 
In the BPT diagram,
our composite  models with $Z$ = 0.6
-- 2.5\,\zsun, the considered range of $\eta$, and a rather small range in $U$
for the AGN (between 0.01 and 0.03) cover the right wing of the seagull quite well.  The reason why the composite
models do not deviate much from the HII region sequence at low
metallicity is clear: it is at high metallicities (solar or higher)
that increasing the hardness of the radiation field has the largest
effect on collisionally excited lines (see Stasi\'nska 2005 for a
discussion of this aspect).  Therefore, panel b of Fig. 4 suggests
that, even if low metallicity AGN existed, they would not  be 
recognized as such in these emission line ratios diagrams. 
As far as we are aware, there is at present no hint on the existence
of low-metallicity Seyfert nuclei. However, is the AGN phenomenon indeed
related to high metallicities or is this belief result of a selection
effect?

In Fig.  5, we show the same kind of
composite models as in Fig. 4, but this time with the metallicity of
the AGN always  fixed at 1.5\,\zsun. We see that composite
models with $Z$ $<$ 0.4\,\zsun\ lie rather close to the pure HII region
sequence, but this time slightly below it. These models are of course
too crude to draw any conclusion on the real effect of a hidden AGN in
a metal poor galaxy. On the other hand, at metallicities $Z$ $\ge$ 
0.4\,\zsun, these composite
models are in good agreement with the observational diagrams.  The
\oiii/\Hb\ vs \sii/\Ha\ diagram is the least well reproduced, but by
playing a bit more with the parameters (radiation field, gas density in the AGN) should
improve the match.

We have thus found a physical and 
quantitative explanation for the distribution of observational points 
in the four usual line-ratio diagnostic diagrams. The only thing that 
we cannot say from these diagrams is whether or not there are AGNs in 
low metallicity galaxies.

Our models suggest that objects along the right wing
differ mainly in the balance between massive stars and AGN ionizing
powers (ie., the mixing parameter $\eta$), with the AGN $U$ acting as
a second parameter (other AGN-related parameters could come into play as well).
Let us compare our sequences of composite models in Fig. 4b with the
Kauffmann et al. line and the Kewley et al. line shown in Fig. 1. We
find that the Kauffmann et al. line corresponds to composite models in which
the AGN contribution to \Hb\ contribution is no more than 3\%.  The
Kewley et al. line is much less restrictive, and allows for an AGN
contribution of roughly 20\%. 

\subsection{The seagull's wings explained!}

One interesting thing to note is that, of the four diagrams presented
here, the only one which shows two  wide open wings is the
\oiii/\Hb\ vs \nii/\Ha. This is actually a consequence of the fact
that, at metallicities larger than about 0.3\,\zsun, N/O increases with
O/H. We know this from observations of galaxies with metallicities between 0.2 and 0.65\,\zsun\  (see e.g. Izotov et al.  2006), where the
measurements of the abundances are obtained from direct, empirical
methods.  For more metal-rich objects, the abundances are generally
derived from empirical methods of statistical value, although in a few
cases abundances can now be derived by direct methods also (which, in
this case, should be examined for possible biases as shown by
Stasi\'nska 2005). The observed trends at metallicities above 0.65\,\zsun
(Pilyugin et al. 2003, Bresolin et al. 2005) are clearly of an increase
of N/O with O/H.  There is a large scatter, though, which can be
explained by the enrichment history of the HII regions (see Pilyugin
et al. 2003).  Therefore, as noted in Sec. 4.2, the HII region
sequence falls towards the right in the \oiii/\Hb vs. \nii/\Ha\
diagram as metallicity increases, while it falls towards the left in
the \oiii/\Hb vs. \oii/\Hb\ diagram. When an AGN is added, heating of
the nuclear region boosts all the optical forbidden lines, which, in
the case of the BPT diagram results in the separation of the right
wing. 

\section{Practical ways to distinguish NSF galaxies and AGN host galaxies}

With our understanding of the classical diagnostic diagrams as 
applied to integrated spectra of galaxies, we can now proceed to 
the main subject of this paper: the classification of galaxies into NSF ones and AGN hosts. 


\subsection{The boundaries between NSF galaxies and AGN hosts in classical emission line diagrams}

It is clear from Fig. 1 that, not only the Kewley et al. line, but
also the Kauffmann et al. line are slightly too ``generous'' in
defining the NSF galaxies region. We propose to define pure NSF
galaxies those that lie to the left of the curves defined by Eqs. 6 and
8,  ``hybrid'' galaxies those that lie between that curve and the
Kauffmann et al. line, and AGN galaxies those that lie to the right of
the Kauffmann et al. line.  As in the case of the Kauffmann et
al. line, there is some degree of subjectivity in defining these
boundaries. However, the curve that we propose is closer to the upper
envelope of the NSF wing in the \oiii/\Hb\ vs \nii/\Ha\ diagram, and
is physically motivated, at least at high values of \oiii/\Hb. The
situation is less clearcut on the high metallicity end.

For an easier use, the curve defined by Eqs. (6) and (7) can  be approximated by 
\begin{eqnarray}
y =
(-30.787+1.1358x+0.27297x^2){ \rm tanh}(5.7409x) \nonumber\\
-31.093, 
\end{eqnarray}
where $y$= log (\oiii/\Hb), and $x$= log (\nii/\Ha).
In the following, we will use this expression for the boundary between pure NSF galaxies and galaxies hosting AGNs. This expression is valid for  log (\nii/\Ha) between -2.0 and -0.4. If \oiii/\Hb\ is not measured we consider that a galaxy is an AGN if  log \nii/\Ha\ $>$ -0.4.

As mentioned in Sect. 5.1, the contribution of the AGN to the \Hb\ emission of galaxies below the Kauffmann et al. line  is at most 3\%. The zone of the BPT diagram between the curve defined by Eq. (11) and the Kauffmann line is very populated: almost 20\% of all the objects appearing in the diagram belong to it. This implies that there is quite a proportion of galaxies that host a very weak AGN in the local Universe.

The  \oiii/\Hb\ vs \sii/\Ha\  and  \oiii/\Hb\ vs \oi/\Ha\ diagrams are
obviously less efficient than the  BPT one to classify galaxies, both because
the dichotomy of the galaxy population is not so clear and because, as shown
above, 
simple photoionization models underpredict the  \sii/\Ha\ and \oi/\Ha\  ratios. In addition, the separation between NSF and AGN galaxies occurs at a value of \oi/\Ha\ where this intensity ratio is difficult to measure. 

The \oiii/\Hb\ vs \oii/\Hb\ diagram is expected to be even worse than
the former two to separate NSF and AGN galaxies. The distribution of
the SDSS galaxies in this plane, the fact that \oii/\Hb\ is sensitive
to reddening  and the
behaviour of the model sequences shown in Fig. 2a do not argue in
favour of its use. Yet, it has been used, when only blue spectra are
available or in the case of redshifted galaxies for which the other
diagnostic lines cannot be observed (e.g. Tresse et al. 1995,
Lamareille et al. 2004).  Using our theoretical borderlines defined by
Eq. (12), we find that, in our sample, 4918 objects are
classified as NSF  galaxies.  in the BPT diagram, 6758
are classified as NSF in the \oiii/\Hb\ vs \oii/\Hb\ diagram, while
only 3504 of those are classified as such in both diagrams. Concerning
AGN galaxies, the corresponding counts are 4499, 3853 and 2216
(here, objects for which \oiii/\Hb\ could not be measured, were classified
as AGN galaxies if \nii/\Ha\ $>$ 0.4 or if \oii/\Hb\ $>$ 0.5).  The
comparison between both classification schemes is thus not as bad as
it might seem from a mere glance at the observational
diagrams. However, it is far from fully satisfactory. Note that, here,
we have considered all the galaxies from the initial sample for which
the relevant emission lines could be measured, irrespective of the
uncertainty in the measured intensities. In a study dealing with the
frequency of AGN with respect to other properties of galaxies, one
should also discuss the question of the uncertainties in the line
ratios, as done for example by Carter et al. (2001).

As demonstrated by Kobulnicky \& Phillips (2003), emission lines
equivalent widths can be used instead of line intensities to estimate
the global metallicities of galaxies. With the same arguments, one can
show that equivalent widths ratios can be used in the same way as line
intensity ratios to distinguish NSF and AGN galaxies, which is
particularly useful in the cases of spectra that are not
well calibrated. The same comments as above apply for
equivalent widths diagrams.

\begin{figure*} 
\centerline{
\includegraphics[scale=0.6]{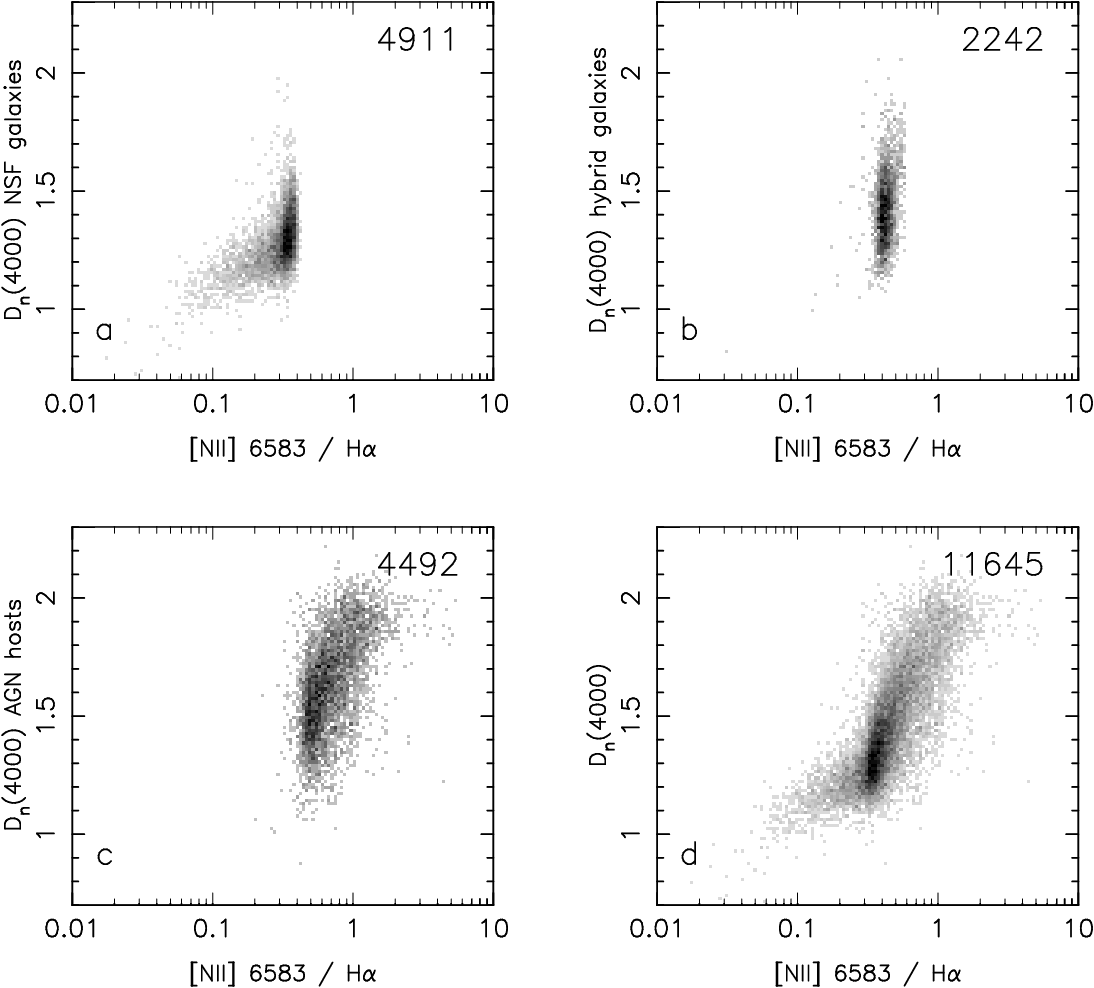}
}
\caption{$D_n(4000)$ versus \nii/\Ha\  for NSF galaxies (panel a), 
hybrid galaxies (panel b), AGN galaxies (panel c) and all our sample 
galaxies (panel d). The total number of 
galaxies in the various plots is indicated in the top right of each 
panel. }
\label{u3}
\end{figure*}

\subsection{A classification based on \nii/\Ha\ only? }

As a matter of fact, since in the BPT diagram the distribution of the
galaxies looks like a flying seagull, one can use the \nii/\Ha\ ratio
\emph {alone} to classify the galaxies. Of course, the \emph
{physical} interpretation of the \nii/\Ha\ ratio would be completely
different for the two wings. For the left wing, it is a measure of the
combination of the metallicity $Z$ and the ionization parameter
$U$. Given the strong correlation between both parameters, as
evidenced by the fact that the left wing is so thin, \nii/\Ha\ can
then be taken as an empirical measure of the gas metallicity. This has
already been mentioned by previous authors for giant HII regions (van
Zee et al. 1998, Denicol{\'o}, Terlevich \& Terlevich 2002, Pettini \& Pagel 2004) and
can be used for \nii/\Ha\ up to 0.3--0.4. Larger values of this ratio
indicate that the galaxies host an AGN. As \nii/\Ha\ increases from
this value upwards, the effect of the AGN on the galaxy spectra
increases and becomes dominant, as can be inferred from Figs. 4b and
5b. However, the right wing of the seagull is rather fuzzy, so that
obviously other parameters enter into play and are not correlated.
 Given our results for the upper limit of the NSF sequence we propose:

\begin{eqnarray}
{ \rm log}  [\rm{N}\,\textsc{ii}] /\Ha \le -0.4 & {\rm NSF},  \nonumber\\
-0.4  < { \rm log} [\rm{N}\,\textsc{ii}] /\Ha \le -0.2 & {\rm Hybrid} , \nonumber\\
 { \rm log} [\rm{N}\,\textsc{ii}] /\Ha > -0.2 & {\rm AGN} 
 \end{eqnarray}

\begin{figure*}
\centerline{
\includegraphics[scale=0.6]{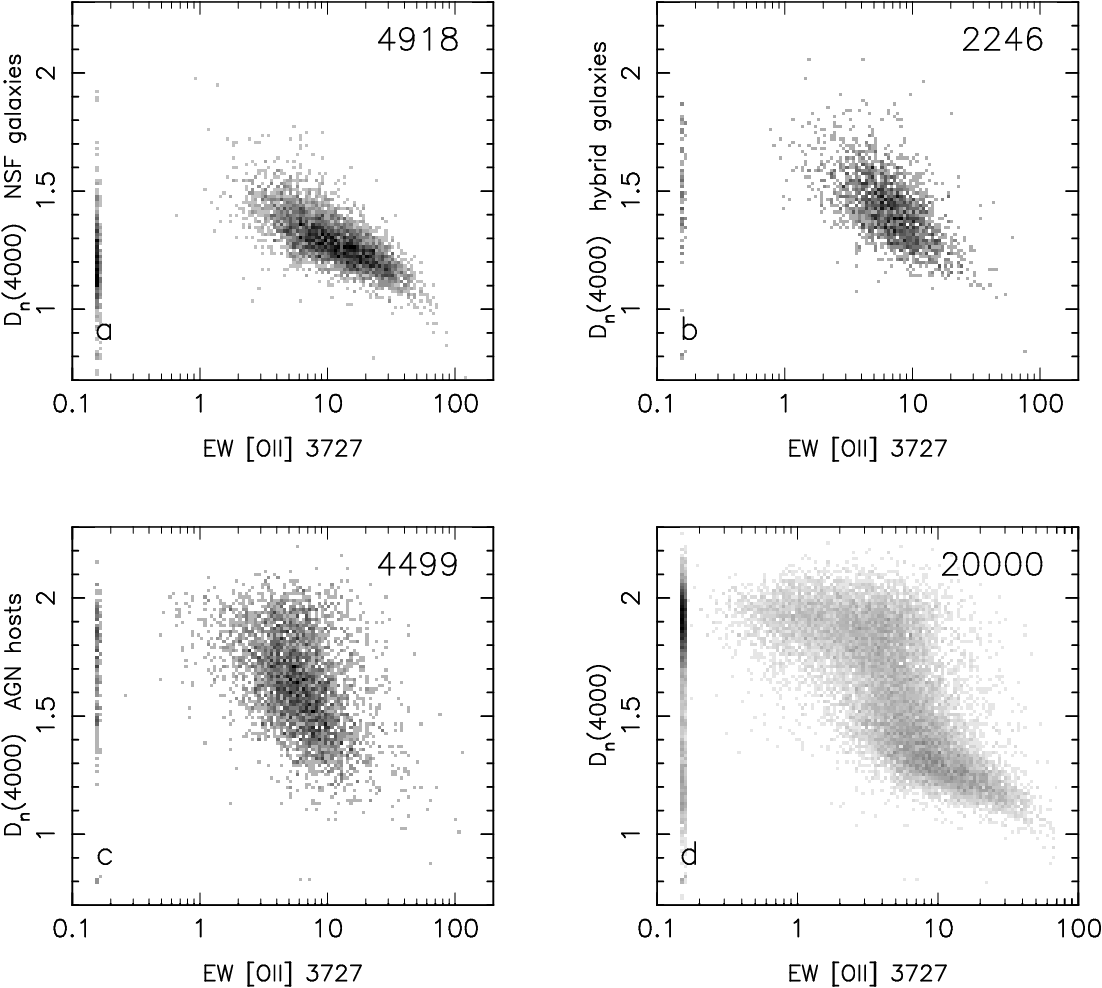}
}
\caption{ $D_n(4000)$  versus  EW\oii\ diagram for NSF galaxies 
(panel a), hybrid galaxies (panel b), AGN galaxies (panel c), and all 
the sample galaxies (panel d). Galaxies with no measurement of 
EW\oii\ are plotted at an abscissa of $\sim$ 0.15, and those with no 
measurement of $D_n(4000)$  are plotted at an ordinate of $\sim$ 0.8.  The total number of 
galaxies (including those with artificially assigned values of the abscissa or of the ordinate) in the various plots is indicated in the top right of each 
panel. }
\label{u3}
\end{figure*}

Being able to distinguish between NSF and AGN galaxies using one
criterion only is very useful since it allows one to study the effect
of any other parameter by a simple 2D-plot. We show an example in
Fig. 6, where we plot the 4000 \AA\ break index, $D_n(4000)$ \footnote
{The break at 4000 \AA\ is defined similarly to Bruzual (1983), who
define D4000 as the ratio between the average value of $F_\nu$ in the
4050--4250 and 3750--3950 \AA\ bands, but using the narrower bands
3850--3950 and 4000--4100 \AA\ introduced by Balogh et al. (1999) to reduce
reddening effects.}, as
a function of \nii/\Ha\ for the NSF galaxies (panel a), the hybrid
galaxies (panel b), the AGN galaxies (panel c) and all the galaxies of
our sample that can be represented in such a plot (panel d). We see
that NSF galaxies tend to have smaller values of $D_n(4000)$ than AGN
galaxies.  There a zone in common, for
$D_n(4000)$ roughly between 1.2 and 1.5. Still, the dichotomy is
important. It indicates that
young galaxies are found only among NSF galaxies, while AGN galaxies
tend to be old and/or metal-rich.  As a
matter of fact, as explained in Sect. 5, the BPT diagram does not
allow one to distinguish low metallicity AGN galaxies (if they exist)
from NSF galaxies, and we already know from our models shown in Sect
5, that objects in the right wing of the seagull are necessarily
metal-rich (or rather: have a metal-rich interstellar medium).

This characteristic behaviour of the galaxies in the $D_n(4000)$
versus \nii/\Ha\ diagram suggests that one could perhaps use the
stellar properties to distinguish NSF and AGN 
galaxies. This is done in the next subsection.

\subsection{A new diagnostic diagram for galaxies at redshifts up to 1.3}

The BPT diagram involves the  \Ha\ and \nii\ lines, meaning  that in a 
survey like the SDSS, which spans the wavelength range 3800 -- 9200 
\AA, it allows one to classify galaxies only up to a redshift of 
about $z = 0.4$. A diagram involving only \oii, \Hb\, and \oiii, which is 
less efficient in separating AGN  from NSF galaxies, as seen 
in the previous sections, could be used for SDSS galaxies up to 
redshifts of about 0.8.

\begin{figure*}
\centerline{
\includegraphics[scale=0.6]{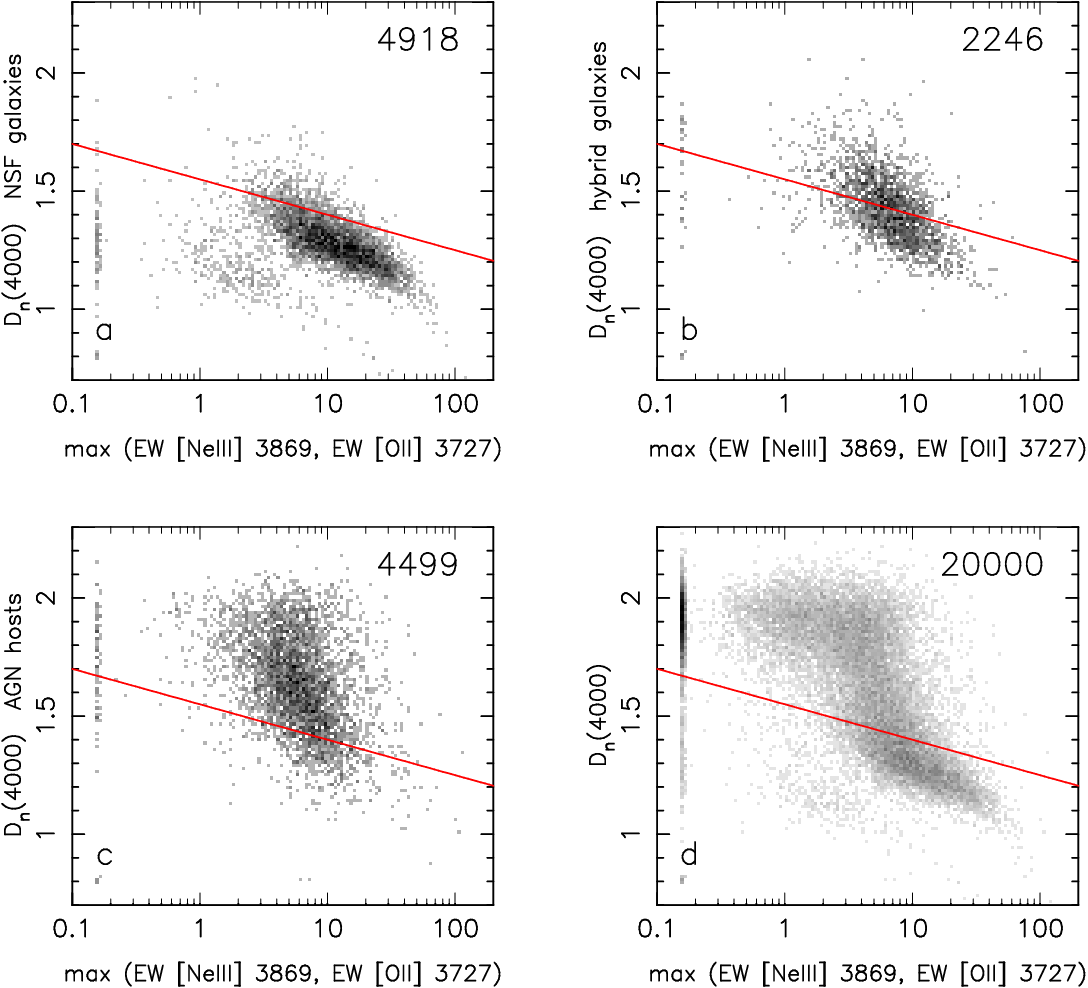}
}
\caption{$D_n(4000)$  versus  max(EW\oii,EW\neiii) for NSF 
galaxies (panel a), hybrid galaxies (panel b), AGN galaxies (panel 
c), and all the sample galaxies (panel d). Galaxies with no 
measurement of EW\oii\ or EW\neiii\ are plotted at an abscissa of 
$\sim$ 0.15, and those with no measurement of $D_n(4000)$  are plotted at an 
ordinate of $\sim$ 0.8. The red line is the adopted boundary between NSF and AGN galaxies, given by Eq. (14). }
\label{u3}
\end{figure*}

If we think that an AGN is a hard non-stellar ionizing source with a
featureless continuum (Koski 1978), the spectrum of a ``pure'' AGN
galaxy should present emission lines without any sign of the presence of
young stars (ages smaller than $10^7$ yr). 
Naturally, young stars can be  associated with an AGN,   in which case they are often
mistaken for a featureless, non-stellar continuum (Cid Fernandes et al
2001, 2004).  However, if we could find a way to segregate at least  ``pure''  AGN
galaxies using their rest-frame blue spectra, it would be a progress
over the present situation, and would allow one extend the classification of galaxies to larger redshifts.

Fig. 6 gives us a clue on how to achieve this goal. The
$D_n(4000)$ index gives a hint on the stellar population.  Large values of   $D_n(4000)$  indicate the presence of a predominantly old stellar population (Cid Fernandes et al. 2005)  and indeed Fig. 6 shows that $D_n(4000)$ tends to be large for AGN hosts and small for NSF galaxies (although there is an important overlap). As for the
emission lines,  their mere presence indicates that ionization is
at work (either due to stars or due to an AGN). A commonly used line in the blue is \oii. Let us then
consider the $D_n(4000)$ versus EW\oii\ diagram. We plot it in
Fig. 7, in 4 panels, which, as in Figs. 6, correspond to NSF, hybrid,
and AGN galaxies as classified by the BPT diagram (panels a, b, and c,
respectively) and to our entire sample (panel d). It is clear that NSF
and AGN galaxies tend to occupy different zones in the plane. In these
figures, we have also represented galaxies with no measurement of
EW\oii\, by plotting them at an abscissa of 0.15, and those with no
measurement of $D_n(4000)$ (there are only a few ones actually), by
plotting them at an ordinate of 0.8.

The \oii\ line may be out of the SDSS spectral range, if the redshift
is very small, or it can be in a noisy zone, close to the limit of the
observable spectral range.  Luckily, the nearby  \Neiii\ line
provides the same kind of information as \oii, namely,  it indicates the presence of ionized 
gas.  In addition, high excitation AGNs may have a \neiii\
line much stronger than the \oii\ line, and may be missed if we use only \oii\ to detect them.  We therefore merge the information provided by  the \oii\  and \neiii\  lines by constructing  a diagram
similar to Fig. 7, but replacing EW\oii\ by max(EW\oii,EW\neiii). This diagram (from now on referred to as the $DEW$ diagram) is shown in Fig. 8. As expected, the number
of galaxies with the relevant measurements is somewhat larger than in
Fig. 7, but the NSF and AGN galaxies continue to occupy different
zones with only a small overlap. Guided by Fig. 8d, we define a new
borderline between NSF and AGN galaxies by the following equation:

\begin{equation}
D_n(4000) = - 0.15 (\log x + 1) + 1.7, 
\end{equation}

\noindent where $x$=max(EW\oii,EW\neiii).

We may now check the correspondence between this new classification
into NSF and AGN galaxies, based on  the $DEW$ diagram, and the classical one based
on the BPT diagram.  We find that 4312 galaxies are classified as NSF
both in the BPT and in the $DEW$ diagram and 3786 are classified as AGN galaxies in
both these diagrams. This is a \emph{ much better} correspondence than
between the BPT and the \oiii/\Hb\ vs \oii/\Hb\ diagram!  We may
visualize this correspondence by plotting the galaxies in the BPT
diagram for the NSF($DEW$) and AGN($DEW$) galaxies separately
(Fig. 9 a and b, respectively). Here, those galaxies without a
measurement of \oiii/\Hb\ are assigned a value of 0.07 for this ratio,
in order to become visible in the plot. We see that this new
classification is in quite good agreement with the one based on the
BPT diagram using the line defined by Eq. (11). There is a
plume of AGN galaxies (according to the BPT diagram) that  are
classified as NSF according to the $DEW$ criterion (eq. 13). This plume
corresponds to the innermost part of the right wing of the seagull,
presumably corresponding to higher values of the ionization
parameter (as suggested by Fig. 4). A more detailed discussion of this is postponed to  a
future paper.

\begin{figure*}
\centerline{
\includegraphics[scale=0.6]{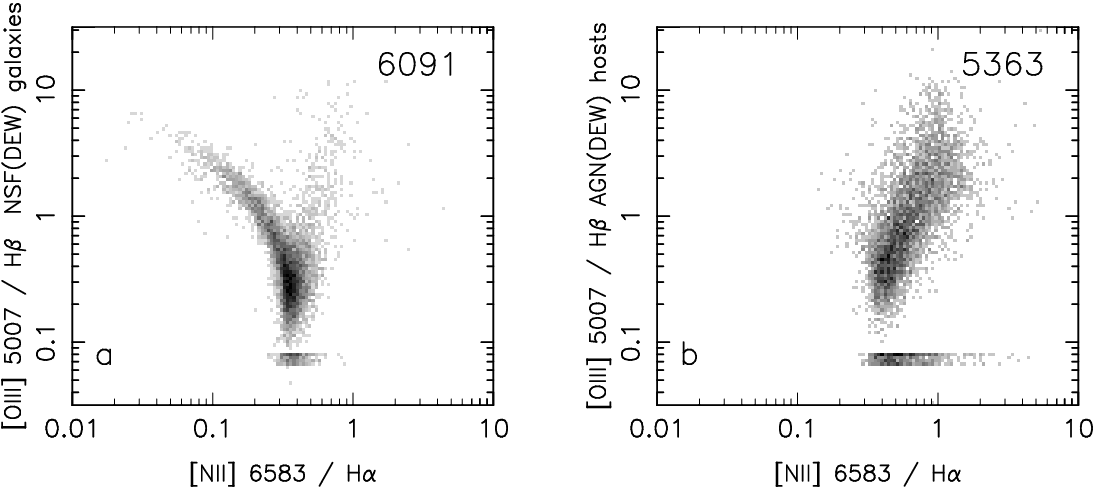}
}
\caption{The BPT diagram for NSF(DEW) galaxies (left) and AGN(DEW) 
galaxies (right). Galaxies without  \oiii/\Hb\ measured are assigned a value of 0.07 for this ratio. The total number of 
galaxies in the various plots is indicated in the top right of each 
panel. }
\label{u3}
\end{figure*}

\section{Summary }

We have considered a sample of 20\,000 galaxies extracted from the
Sloan Digital Sky Survey and constituting a magnitude-limited
sample.  We have applied the spectral synthesis technique
described in previous papers in this series to the spectra of these
galaxies in order to properly subtract the starlight and obtain a pure
nebular spectrum. The emission line intensities have been measured
with our automated procedure.  These data have been used to revisit
the classical diagrams that are used to distinguish normal star
forming galaxies from galaxies hosting an AGN, and to propose new
diagrams.

We first analyzed the four classical emission line ratio diagrams:
\oiii/\Hb\ vs \oii/\Hb, \oiii/\Hb\ vs \nii/\Ha\ (the BPT diagram),
\oiii/\Hb\ vs \sii/\Ha, and \oiii/\Hb\ vs \oi/\Ha. From a purely
observational point of view, the BPT diagram is the one which best
distinguishes two categories of galaxies, as it distributes the
galaxies in two wings which look like the wings of a seagull. The left
wing, identified with the sequence of normal star forming galaxies, is
very narrow. The right wing, which appeared clearly for the first time
in the paper by Kauffmann et al. (2003) also based on SDSS galaxies,
is constituted of galaxies hosting an AGN. We have computed a series
of photoionization models, using as an input the spectral energy distributions from evolutionary stellar population synthesis. We used the population synthesis code Starburst 99 (Leitherer et al. 1999) in the version which incorporates  the most elaborated
stellar atmospheres for the massive stars (Smith et al. 2002). Our photoionization models confirm this
interpretation and allow us to draw physically based divisory lines in
all the four classical diagrams. However, the models are too schematic to
reproduce the observed \sii/\Ha\ and \oi/\Ha\ line ratios
correctly. Therefore, the model sequence that best divides NSF and AGN
galaxies in the \oiii/\Hb\ vs \oii/\Hb\ or \oiii/\Hb\ vs \nii/\Ha\
diagrams cannot be safely used to distinguish NSF and AGN galaxies in
the \oiii/\Hb\ vs \sii/\Ha, and \oiii/\Hb\ vs \oi/\Ha diagrams. 

We propose the following divisory line between NSF galaxies and AGN hosts
 in the BPT diagram:
\begin{eqnarray}
y =
(-30.787+1.1358x+0.27297x^2){ \rm tanh}(5.7409x) \nonumber\\
-31.093, 
\end{eqnarray}
where $y$= log (\oiii/\Hb), and $x$= log (\nii/\Ha), replaced by log (\nii/\Ha)=-0.4 if \oiii/\Hb\ is not available. 
This line is actually  close to the line drawn empirically by Kauffmann et
al. to distinguish NSF galaxies from AGN hosts. We found that the Kauffmann et
al. line  includes among NSF galaxies objects that have an AGN contribution to
\Hb\ of up to 3\%. 
Thus, depending on the problem one is interested in, one may want to use either the Kauffmann. et al line, or the line we propose in this paper, in order to segregate NSF galaxies from AGN hosts. The Kewley line is much less restrictive, and  allows for an AGN contribution of roughly 20\%.

Since the BPT diagram is very populated between the line defined by
Eq. (11) and the Kauffmann line (it contains about 11\% of the
galaxies in our sample, including passive galaxies), this means that
the local Universe contains a fair proportion of galaxies with very
low level nuclear activity, in agreement with the statistics from
observations of galactic nuclei eg., Ho, Fillipenko \& Sargent
(1997).

We  point out that emission line ratio diagrams are not efficient in detecting the presence of an AGN in low metallicity galaxies, if such
cases exist.

We have shown that  a classification into NSF and AGN galaxies using only \nii/\Ha\ 
 is feasible and useful.

Finally, we propose a new classification diagram (named the $DEW$ diagram), which uses 
$D_n(4000)$  vs  max(EW\oii,EW\neiii).
This classification has many advantages:
\begin{itemize}
\item It can be used at much  larger redshifts than the previous emission line classifications. With SDSS spectra, it can be applied to galaxies with redshifts up to $z = 1.3$.
\item It requires only
a small range in wavelengths, so it can also be used at even larger
redshifts in suitable windows in the near infra-red. 
\item It  can be used without a stellar synthesis analysis to subtract the stars.
\item It allows one to see \emph{all} the galaxies in the same
diagram, including passive galaxies (the definition of passive assumes
a certain detection limit of emission lines).  Hence, all galaxies can be
classified.
\end{itemize}

This method has drawbacks too:
\begin{itemize}
\item  It is not exactly equivalent to the usual BPT classification. But 
does it matters?
\item  Old galaxies with a recent starburst ($< 10^7$ yr) will be mistaken for AGN hosts.
\item The borderline between NSF and AGN galaxies is somewhat ``porous'' (but this is the case of
almost any frontier). Note that in the BPT diagram, the 
borderline is also not very well defined at the low excitation end.
\end{itemize}

We note that our proposed classification in the $DEW$ diagram is actually more compatible with that based on the  \oiii/\Hb\ vs 
\nii/\Ha\ diagram (when using our boundary line) than a classification based on the \oiii/\Hb\ vs \oii/\Hb\ diagram which is used in some papers.

With this new classification scheme at hand, it will be possible to investigate the evolution of AGN galaxy populations in a much larger redshift range than has been done so far, and on  firmer grounds.

\section*{Acknowledgments}

We gratefully acknowledge financial support from CNPq, FAPESP and the
France-Brazil PICS program.
All the authors wish to thank the team of the
Sloan Digital Sky Survey (SDSS) for their dedication to a project
which has made the present work possible.

The Sloan Digital Sky Survey is a joint project of The University of
Chicago, Fermilab, the Institute for Advanced Study, the Japan Participation
Group, the Johns Hopkins University, the Los Alamos National Laboratory, the
Max-Planck-Institute for Astronomy (MPIA), the Max-Planck-Institute for
Astrophysics (MPA), New Mexico State University, Princeton University, the
United States Naval Observatory, and the University of Washington.
Funding for the project has been provided by the Alfred P. Sloan Foundation,
the Participating Institutions, the National Aeronautics and Space
Administration, the National Science Foundation, the U.S. Department of Energy,
the Japanese Monbukagakusho, and the Max Planck Society.

{}
\end{document}